%% file: main.tex
\documentclass[conference]{IEEEtran}
\IEEEoverridecommandlockouts

\usepackage{cite}
\usepackage{amsmath,amssymb,amsfonts}
\usepackage{algorithmic}
\usepackage{graphicx}
\usepackage{textcomp}
\usepackage{xcolor}
\usepackage[acronym,ucmark,translate]{glossaries}

\usepackage{url}
\usepackage{booktabs}
\usepackage{float}

\def\BibTeX{{\rm B\kern-.05em{\sc i\kern-.025em b}\kern-.08em
    T\kern-.1667em\lower.7ex\hbox{E}\kern-.125emX}}

\input{glossary.tex}

\newglossary[slg]{symbols}{syi}{syg}{List of symbols}
\makeglossaries
\begin{document}

\title{RTL Fault Injection of a Deployed Graph Neural Network Trigger for Belle~II}

\author{\IEEEauthorblockN{
Georgios Sotiropoulos\IEEEauthorrefmark{1},
Marc Neu\IEEEauthorrefmark{1},
Tanja Harbaum\IEEEauthorrefmark{1},
Torben Ferber\IEEEauthorrefmark{1},
Jürgen Becker\IEEEauthorrefmark{1}
}

\IEEEauthorblockA{
\IEEEauthorrefmark{1}Karlsruhe Institute of Technology (KIT) \\
Email: \{georgios.sotiropoulos, marc.neu, harbaum, torben.ferber, becker\}@kit.edu
}
}

\maketitle

\begin{abstract}
    As particle physics detectors grow in scale, High Energy Physics experiments must process ever-increasing data volumes. Level-1 trigger systems, implemented on Field-Programmable Gate Arrays and increasingly using neural-network algorithms, filter this data in real time. However, their proximity to the interaction point exposes them to radiation, which can corrupt outputs, stall processing pipelines, or damage hardware, with significant financial and scientific consequences. In this work, we present the first Register Transfer Level fault-injection study of a deployed Level-1 hardware neural-network trigger, GNN-ETM in the Belle~II trigger system. We target three failure modes most consequential to a real-time trigger pipeline: deadlocks, timeouts, and packet-integrity violations. Through two complementary campaigns, we inject $1\,442\,840$ Single-Event Upsets across $211\,245$ signals. We find a monitoring asymmetry in the existing verification infrastructure and propose inter-stage liveness monitoring as a more accurate alternative to output-only observation, showing that Mean Time To Failure estimates from the two approaches differ by up to 78.7\%. The resulting per-stage data identifies the highest-priority hardening targets.
\end{abstract}

\input{sections/1_introduction}
\input{sections/2_related_work}
\input{sections/3_system_architecture}
\input{sections/4_fault_injection_methodology}
\input{sections/5_results}
\input{sections/6_conclusion}

\section*{Acknowledgment}
{This work is funded by the German Federal Ministry of Research, Technology and Space (BMFTR) in the framework of design tools for sovereign chip development with open source (DE:Sign DI-EDAI, grant number 16ME0990K) and by the DEEP consortium (05D25VK1) in the ErUM-Data action plan.}

\bibliographystyle{IEEEtran}
\bibliography{IEEEabrv,biblio}

\end{document}

%% file: glossary.tex
\newacronym{AI}{AI}{Artificial Intelligence}
\newacronym{ML}{ML}{Machine Learning}
\newacronym{FPGA}{FPGA}{Field-Programmable Gate Array}
\newacronym{TMR}{TMR}{Triple Modular Redundancy}
\newacronym{RTL}{RTL}{Register Transfer Level}
\newacronym{NN}{NN}{Neural Network}
\newacronym{GNN}{GNN}{Graph Neural Network}
\newacronym{CNN}{CNN}{Convolutional Neural Network}
\newacronym{HLS}{HLS}{High-Level Synthesis}
\newacronym{HEP}{HEP}{High Energy Physics}
\newacronym{L1}{L1}{Level-1}
\newacronym{FI}{FI}{Fault Injection}
\newacronym{ECL}{ECL}{Electromagnetic Calorimeter}
\newacronym{TC}{TC}{Trigger Cell}
\newacronym{GPU}{GPU}{Graph Processing Unit}
\newacronym{PE}{PE}{Processing Element}
\newacronym{MTTF}{MTTF}{Mean Time To Failure}
\newacronym{SRAM}{SRAM}{Static Random-Access Memory}
\newacronym{LUT}{LUT}{Look Up Table}
\newacronym{GDL}{GDL}{Global Decision Logic}
\newacronym{DAQ}{DAQ}{Data Acquisition}
\newacronym{DSP}{DSP}{Digital Signal Processor}
\newacronym{AXI}{AXI}{Advanced eXtensible Interface}
\newacronym{MAC}{MAC}{Media Access Control}
\newacronym{SDC}{SDC}{Silent Data Corruption}
\newacronym{DUE}{DUE}{Detected Unrecoverable Error}
\newacronym{SEU}{SEU}{Single Event Upset}
\newacronym{DUT}{DUT}{Device Under Test}

%% file: sections/1_introduction.tex
\section{Introduction}
\label{sec:introduction}

Modern particle physics detectors have grown significantly and are capable of generating extremely large volumes of data. To reduce the amount of data for offline storage, \gls{HEP} experiments employ trigger systems, which are responsible for filtering, in real time, the data to be saved for further analyses. Because trigger applications must satisfy strict latency and throughput constraints, they are typically implemented on \glspl{FPGA}.

In recent years, \gls{NN} models have been increasingly integrated into trigger systems for a variety of tasks in several experiments, such as Belle~II \cite{Bahr:2024dzg,Liu:2026iup,gnn_etm} and CERN Compact Muon Solenoid (CMS) detector \cite{zipper:2024,Gandrakota_2024}. However, as these \gls{NN}-based trigger systems are deployed close to the interaction point, they are exposed to radiation, caused by particle collisions in the detector \cite{Wirthlin_2015}.

\begin{figure}[!t]
    \centering
    \includegraphics[width=\columnwidth]{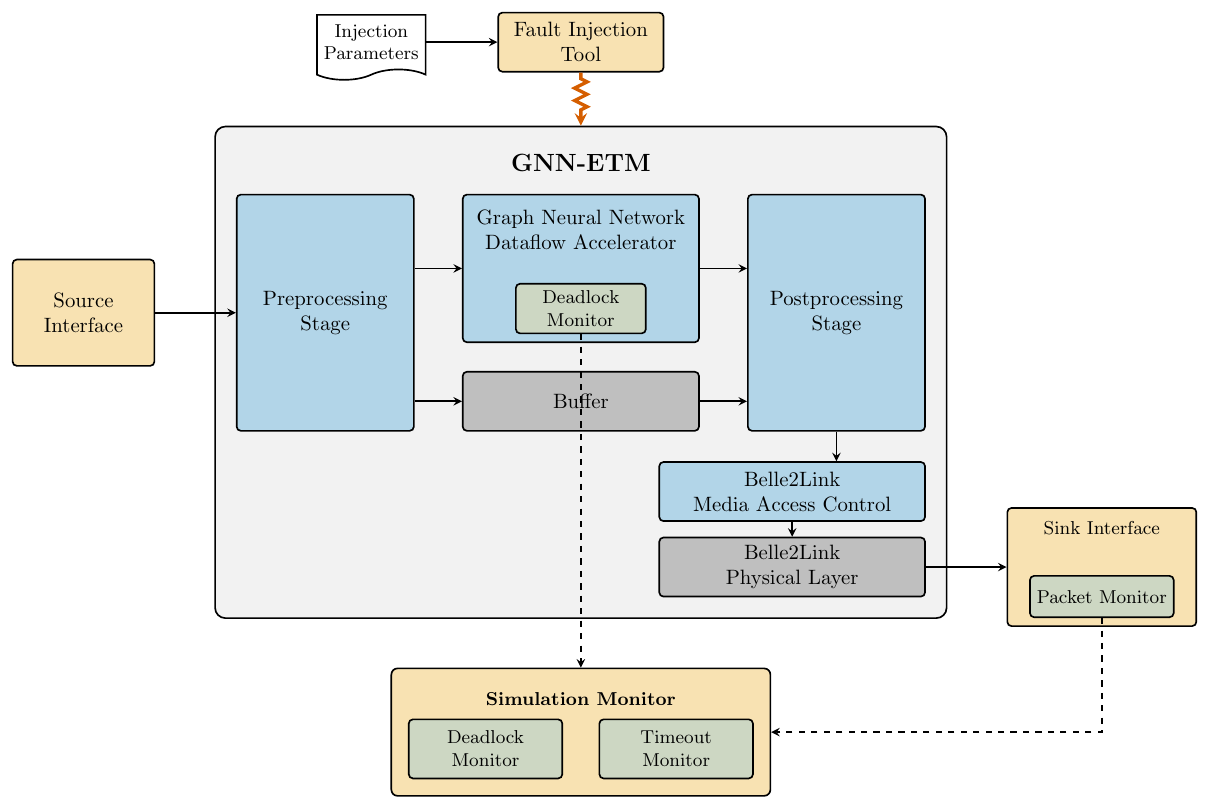}
    \caption{Overview of the RTL fault injection testbench used to evaluate the reliability of GNN-ETM.}
    \label{fig:testbench_arch}
\end{figure}

Radiation-induced faults in hardware can lead to corrupted output data, stalled processing pipelines or even cause permanent hardware damage~\cite{Dodd_2010}. Given the scale of investment, failures in the trigger logic that cause downtime can result in substantial financial and scientific costs.

Prior fault-injection studies on \gls{NN} accelerators have concentrated on \gls{SDC}, while \glspl{DUE} (hangs, deadlocks, timeouts) remain understudied~\cite{ahmadilivani_2024}. For a real-time trigger, \glspl{DUE} and packet-integrity violations are the most consequential failures. Existing studies also evaluate benchmark workloads on representative accelerators rather than full end-to-end deployed systems.

\input{figures/tab_related_work}

A timely and realistic case study for a fault tolerance analysis is the Belle~II trigger system, presented in~\cite{gnn_etm}, which is an \gls{FPGA}-based \gls{GNN} dataflow accelerator for calorimeter clustering and trigger-bit generation. Its scale and complexity render traditional mitigation techniques such as \gls{TMR} impractical, since the associated area overhead would prevent the design from fitting on a single FPGA. Moreover, the latency overhead of using multiple \glspl{FPGA} would violate the strict timing constraints.

In this paper, we investigate how a complex hardware trigger can be affected by \glspl{SEU}. Our contributions are as follows.
\begin{itemize}
    \item We present, to the best of our knowledge, the first \gls{RTL} \gls{FI} study of a deployed \gls{L1} hardware trigger running a \gls{NN} workload, targeting GNN-ETM in the Belle~II trigger system.
    \item We characterise three types of failures most consequential to a real-time trigger pipeline: deadlocks, timeouts, and packet-integrity violations. Packet-integrity violations sit between the established \gls{SDC} and \gls{DUE} classes in terms of criticality.
    \item We propose inter-stage liveness monitoring as a more accurate alternative to output-only observation, showing that \gls{MTTF} estimates from the two approaches differ by up to 78.7\%. With the resulting data we identify the highest-priority hardening targets.
\end{itemize}

%% file: figures/tab_related_work.tex
\begin{table*}[!t]
    \centering
    \caption{Comparison of fault-injection studies of FPGA-based accelerators and the present work.}
    \label{tab:related_work}
    \setlength{\tabcolsep}{6pt}
    \footnotesize
    \begin{tabular}{lllll}
        \toprule
        \textbf{Work} & \textbf{Injection layer} & \textbf{Fault type} & \textbf{Target DUT} & \textbf{Failure taxonomy} \\
        \midrule
        \cite{Xu_2021}                          & Config. memory + BRAM      &  Stuck-At & 2D PE Array               & Accuracy + system exception \\
        \cite{Tonfat_2017}                      & Config. memory             & SEU      & HLS matmul benchmark       & SDC + timeout \\
        \cite{Jonckers_2024, Jonckers_2026}     & RTL flip-flops             &  SEU       & Systolic-array DNN         & SDC + misclassification \\
        \textbf{This work}                      & \textbf{RTL signals}       & \textbf{SEU} & \textbf{Deployed L1-trigger GNN accelerator} & \textbf{Deadlock / timeout / packet integrity} \\
        \bottomrule
    \end{tabular}
\end{table*}

%% file: sections/2_related_work.tex
\section{Related Work}
\label{sec:related_work}

Ahmadilivani et al.~\cite{ahmadilivani_2024} systematically review 139 papers on \gls{NN} reliability analysis, identifying three evaluation approaches (fault injection, analytical, hybrid). The survey finds that prevailing methodology focuses on \gls{SDC} while \glspl{DUE} (hangs, crashes, deadlocks) remain understudied, and identifies the control part of hardware \gls{NN} accelerators as an understudied area.

Rech~\cite{Rech_2024} reviews radiation effects in \gls{NN} systems, covering fault propagation and hardening strategies, with a focus on \glspl{GPU} and \glspl{FPGA}. The survey notes that \gls{RTL} fault injection provides full observability at the cost of longer simulation times, and that certain execution-flow failures are invisible to software-level fault injection.

Xu et al.~\cite{Xu_2021} evaluated the reliability of a representative 2D \gls{PE} array accelerator, by injecting stuck-at faults. The design is deployed on an AMD Zynq-7000 SoC platform, running four representative \glspl{NN} under varying error rates. The authors observe that system exceptions (IO stalls, timeouts and early terminations) dominate the failures that occurred, compared to accuracy degradation. Applying \gls{TMR} to the critical modules reduces the number of failures by a small amount, which is attributed to inaccessible to designers logic supplied by the vendor. However, two methodological limitations can be identified. First, faults are injected across the entire design, targeting only the memory, making it difficult to attribute failures to specific components. Additionally, system exceptions are observed only from the host processor perspective, which may delay their detection, consequently skewing \gls{MTTF} estimates

Tonfat et al.~\cite{Tonfat_2017} distinguish \gls{SDC} and timeout outcomes in a configuration-memory \gls{FI} campaign of \gls{HLS}-generated matrix-multiplication variants on an AMD~Artix-7. They show that control-flow-heavy designs fail through timeouts while aggressively pipelined data-flow variants fail mostly as single-output \glspl{SDC}. Their emulation-based injection acts exclusively on configuration memory bits.

A closely related line of work, with respect to the injection approach, is provided by the recent studies of Jonckers et al.~\cite{Jonckers_2024, Jonckers_2026}, which apply \gls{RTL} \gls{FI}, to a \gls{NN} accelerator, covering both the systolic-array core and the post-processing chain. In their 2024 study~\cite{Jonckers_2024}, the authors characterize the fault-propagation probability and fault magnitude of different register groups under constrained-random inputs, identifying the 32-bit systolic-array and accumulator registers as high-value hardening targets under Amdahl's law. In a follow-up study~\cite{Jonckers_2026}, they extend the analysis to three image-classification \gls{NN} workloads and distinguish output faults into non-critical perturbations and misclassifications, showing that both model redundancy and training accuracy, can shift the critical-fault rate by roughly an order of magnitude. It is worth noting that the failure taxonomy does not consider \glspl{DUE}, the faults injected target only the Flip-Flops and a single fault is injected per inference run, which doesn't capture the accumulation of faults during real world operation.

In this work, we study how faults lead to \glspl{DUE} in a \gls{GNN} accelerator deployed in the Belle~II trigger system, as summarised in Table~\ref{tab:related_work}.

%% file: sections/3_system_architecture.tex
\section{System Architecture}
\label{sec:system_architecture}

\subsection{GNN-ETM Design}

Belle~II at SuperKEKB uses a hardware \gls{L1} trigger composed of dedicated subtriggers, one per participating subdetector, operating under a hard real-time deadline of $4.4\,\mu$s. GNN-ETM~\cite{gnn_etm} is the \gls{L1} trigger module for the \gls{ECL} subdetector. It performs real-time cluster reconstruction from \glspl{TC} and generates the corresponding \gls{L1} trigger bits. The version used in this study differs from the original in two relevant ways: it uses a single GravNet block (vs.\ two) and includes the \gls{L1} trigger-bit generation logic. The full network architecture --- CaloClusterNet, GravNet message passing, object-condensation training, and condensation-point selection --- is described in~\cite{gnn_etm}.

\begin{figure}[t]
    \centering
    \includegraphics[width=\columnwidth]{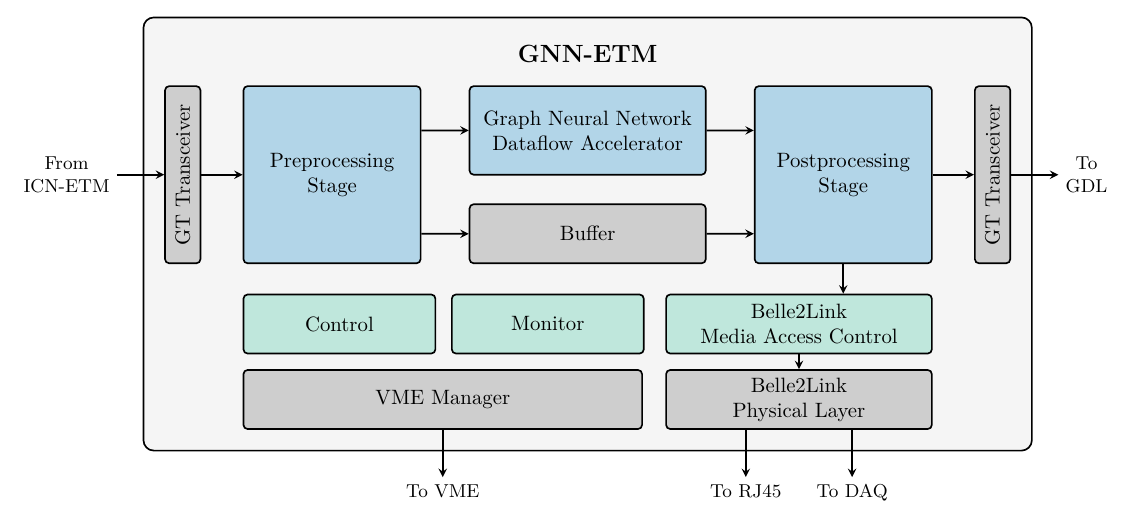}
    \caption{Architecture overview of GNN-ETM, comprising the Preprocessing Stage, GNN Dataflow Accelerator (running CaloClusterNet inference), Postprocessing Stage, and Belle2Link \gls{MAC}.}
    \label{fig:gnn_etm_arch}
\end{figure}

GNN-ETM is implemented as a pipelined dataflow accelerator with four sub-components, shown in Fig.~\ref{fig:gnn_etm_arch} and with signal counts summarised in Table~\ref{tab:subcomponent_nodes}. The Preprocessing Stage normalises and scales the incoming \gls{TC} feature vectors and contributes $385\,$ns to the end-to-end latency. The GNN Dataflow Accelerator runs CaloClusterNet inference. It dominates both latency ($821\,$ns) and resource usage, consuming all available \glspl{DSP} and approximately half the \glspl{LUT}. The Postprocessing Stage performs the post-inference condensation-point selection, selecting cluster centroids from the GNN output, and contributes $165\,$ns. The Belle2Link \gls{MAC} provides the Belle2Link protocol interface, multi-stream synchronisation, \gls{DAQ} buffering, and arbitration, contributing the remaining $475\,$ns. The total end-to-end latency is $1846\,$ns. The system is implemented on a Universal Trigger Board (UT4) equipped with an AMD UltraScale XCVU190 \gls{FPGA}.

The heterogeneous source code (\gls{HLS} with AMD Vitis for the GNN Dataflow Accelerator, Chisel for the other three sub-components) is compiled to Verilog and integrated by FuseSoC into a single flat-netlist \gls{RTL} design, which is the input to our \gls{FI} campaigns.

\subsection{Baseline Failure Detection}
\label{sec:baseline_detection}

The existing failure-detection setup uses two mechanisms, both active only during simulation. An \gls{HLS}-generated watchdog inside the GNN Dataflow Accelerator flags pipeline deadlocks within the GNN block. The Belle2LinkChecker, in the cocotb testbench at the design output, validates protocol conformance of the b2link byte stream and flags packet-integrity violations, in a similar way to the design-output observation approach of~\cite{Xu_2021}. Together, these mechanisms cover the GNN block and the design output. Failures originating elsewhere in the pipeline are observed only indirectly. The cocotb testbench also instruments the inter-stage \gls{AXI}-Stream interfaces with transaction monitors for functional verification, which are not part of the failure-detection setup. GNN-ETM itself currently includes no fault-tolerance mechanisms, such as configuration-memory scrubbing or module-level redundancy. This setup defines the \emph{baseline detection} view used in the comparison study of Section~\ref{sec:results}.

\begin{table}[t]
    \centering
    \caption{Number of signals in GNN-ETM and its four sub-components.}
    \label{tab:subcomponent_nodes}
    \begin{tabular}{lrr}
        \toprule
        \textbf{Sub-component} & \textbf{\# Signals} & \textbf{\% of Design} \\
        \midrule
        Full Design                     & 211\,245 & 100.00 \\
        Preprocessing Stage             &  47\,902 &  22.68 \\
        GNN Dataflow Accelerator        & 116\,158 &  54.99 \\
        Postprocessing Stage            &   6\,010 &   2.84 \\
        Belle2Link \gls{MAC}            &  40\,982 &  19.40 \\
        \bottomrule
        \addlinespace[1pt]
        \multicolumn{3}{l}{\scriptsize The remaining $0.09\%$ are small reset modules, excluded as obvious mitigation targets.} \\
    \end{tabular}
\end{table}

%% file: sections/4_fault_injection_methodology.tex
\section{Fault Injection Methodology}
\label{sec:fault_injection_methodology}

\subsection{Simulation Setup and Fault Model}

We analyse our \gls{DUT}, GNN-ETM, by injecting faults through \gls{RTL} simulation. Compared to alternatives such as beam testing and emulation, this enables more flexibility in choosing specific areas for \gls{FI} and a target-platform-agnostic analysis~\cite{Rech_2024}. An overview of the resulting fault-injection testbench is shown in Fig.~\ref{fig:testbench_arch}.

Each simulation run takes approximately $250\,000$\,ns of simulated time with multiple faults injected per run. The \gls{FI} follows a Poisson process, which is the model used for \gls{SEU} ground testing~\cite{Faure_2005} and in-detector \gls{FPGA} operation~\cite{Scialdone_2022}. The injection rate is set to one fault per $1\,846$\,ns, the design's end-to-end latency. This is several orders of magnitude higher than the real radiation rate at the detector, which is why fault-injection studies normally inject a single fault per run. An exhaustive single-fault-per-run sweep across all $211\,245$ signals would have required approximately 488 days to complete even with parallelism, so we trade absolute fidelity for simulation feasibility. The campaign therefore quantifies relative sub-component reliability rather than absolute in-detector failure rates.

We focus on faults that cause deadlocks, timeouts, or packet-integrity violations. A \emph{deadlock} is a state in which the design, or one of its sub-components, stops making forward progress. A \emph{timeout} is flagged when a simulation run exceeds the expected simulated time without completing. A \emph{packet-integrity violation} corresponds to an output stream whose structure no longer conforms to the Belle2Link protocol and is therefore unreadable by the downstream \gls{DAQ}. A simulation run is classified as failed if any of these is observed, and as passed otherwise.

Simulations ran in ModelSim~\cite{modelsim:2023:4} with the cocotb Python verification framework~\cite{cocotb:2025}. Faults were injected with the open-source cocotb fault injection tool~\cite{cocotb_fault_injection} developed at CERN. The input stimulus consists of randomly generated \gls{TC} data, with a fresh random seed per run, since the targeted failure modes are caused by faults in control signals rather than by the data being processed. All experiments ran on a server equipped with an AMD EPYC 7702P 64-core processor and 503\,GB of RAM. Each fault is a single-bit \gls{SEU} injected at signal level: a single bit of the targeted signal is flipped and remains flipped until the next normal write to that signal. Multi-bit upsets, relevant at advanced FPGA nodes~\cite{Quinn_2005}, are not injected here but would propagate to the \gls{AXI}-Stream interfaces identically and would be detected by the same inter-stage liveness monitors. We treat all \gls{RTL} signals uniformly and do not report their storage-vs-combinational breakdown, since separating them requires synthesis that is either unreliable for the design's HLS output and locked vendor IP (open-source flows) or tied to a specific toolchain (vendor flows). Faults target signals from both the full design and each sub-component individually. Injecting faults at the sub-component level is necessary since, when a failure occurs, it is otherwise difficult to trace which fault was responsible. For each target, we run enough simulations to inject at least as many faults as the number of signals it contains. The per-target run and fault counts are reported in Table~\ref{tab:campaign_summary}.

\subsection{Two Injection Campaigns}

\label{sec:two_campaigns}
To analyze GNN-ETM, we ran two separate \gls{FI} campaigns. The \emph{proportional-rate} campaign mimics a uniform radiation environment: each sub-component is injected at a rate proportional to its share of the design's signals. The \emph{fixed-rate} campaign instead injects every target at the same absolute rate as the full design, so that \gls{MTTF} is directly comparable across sub-components.

The proportional rate follows from the standard relationship between particle flux and device sensitivity. The particle flux $\Phi$ is the number of particles crossing a unit area per unit time. The cross-section $\sigma$ of a device is the effective sensitive area within which an incident particle produces a fault~\cite{Rech_2024}. From these two quantities, the expected fault rate of the device, in faults per unit time, is
\begin{equation}
    \lambda = \Phi \cdot \sigma.
\end{equation}
We assume that the sensitive area is distributed uniformly across the design, meaning that every signal contributes an equal share of the total cross-section. Per-signal cross-sections cannot be characterised at \gls{RTL}, since they depend on the specific FPGA mapping and require beam-irradiation measurements, making the uniform cross-section assumption necessary for a platform-agnostic study. Therefore, the cross-section of any portion of the device is calculated as $\sigma = \sigma_s \cdot N$, where $\sigma_s$ is the per-signal cross-section and $N$ the number of signals. Having already fixed the injection rate of the full design to $\lambda_{\text{full}}$, corresponding to one fault per $1\,846$\,ns (the end-to-end latency of the design), the same particle flux combined with the uniform-$\sigma_s$ assumption yields the adjusted per-component rate
\begin{equation}
    \lambda_{\text{comp}} = \lambda_{\text{full}} \cdot \frac{N_{\text{comp}}}{N_{\text{full}}},
\end{equation}
with $N_{\text{comp}}$ and $N_{\text{full}}$ taken from Table~\ref{tab:subcomponent_nodes}. The \emph{proportional-rate} campaign applies $\lambda_{\text{comp}}$ to each sub-component in isolation, with the rest of the design left unfaulted. The relative rates across sub-components therefore mirror those of a uniform radiation environment, while absolute rates are anchored to the full-design rate chosen for simulation feasibility and are higher than the real-world equivalent.

The proportional-rate campaign captures relative exposure but mixes two effects: how often a sub-component is hit and how sensitive it is per fault. Small components like the Postprocessing Stage ($6\,010$ signals) see only a handful of faults per run, which inflates their \gls{MTTF} and makes per-component sensitivity hard to compare directly. The fixed-rate campaign complements it by equalising exposure across targets, isolating per-fault sensitivity.

\subsection{Inter-Stage Liveness Analysis}
\label{sec:inter_stage_liveness}

The simulations themselves run under the baseline detection setup of Section~\ref{sec:baseline_detection}. We additionally re-analyse the same simulation logs with an inter-stage liveness check, deriving four per-stage deadlock signatures from the cocotb transaction monitors at the inter-stage \gls{AXI}-Stream interfaces. Section~\ref{sec:results} compares the two views.

The four monitored interfaces, in pipeline order, are:
\begin{itemize}
    \item \texttt{Pre.\ header}: the path from preprocessing to the Belle2Link header.
    \item \texttt{GNN DF output}: the output of CaloClusterNet inference.
    \item \texttt{Post.\ CPS}: the path from postprocessing to the Belle2Link CPS.
    \item \texttt{Post.\ GDL}: the path from postprocessing to the \gls{GDL}.
\end{itemize}
During normal operation each stream emits a transaction roughly every $125\,$ns. A monitored stream is declared deadlocked if it remains silent for more than $2\,\mu$s after the $1\,846$\,ns end-to-end latency has elapsed, well above the normal interval, to avoid false positives. A tighter threshold would lower \gls{MTTF} estimates at the cost of more false positives from natural backpressure, but the coverage gain over the baseline holds at any reasonable threshold. An equivalent live monitor would catch the same events with the same detection latency: the analysis only inspects \gls{AXI}-Stream handshake events that a live monitor would also observe in real time, and detection is in both cases bounded by the $2\,\mu$s silence window.

When classifying each run, inter-stage deadlock signatures take priority. Baseline mechanisms (HLS watchdog, Belle2LinkChecker) classify the run if no signature fires.

%% file: sections/5_results.tex
\section{Experimental Results}
\label{sec:results}

\input{figures/tab_campaign_summary.tex}

\input{figures/tab_fi_results_T.tex}

The two \gls{FI} campaigns inject a total of $1\,442\,840$ \glspl{SEU} across the full GNN-ETM design and its four sub-components, as shown in Table~\ref{tab:campaign_summary}. The Full Design row is identical between the two campaigns, so its faults are counted only once in this total. Table~\ref{tab:fi_results} summarises the outcomes of both campaigns under the baseline and inter-stage monitoring approaches: total runs, the per-class failure distribution as a percentage of total runs, the \gls{MTTF} in microseconds of simulated time, and the percentage \gls{MTTF} deviation between them. Each column of Table~\ref{tab:fi_results} corresponds to a separate batch of simulation runs with faults injected at a different target (full design, or only one sub-component), so percentages should be read within a column rather than compared across columns.

The Margin of error row in Table~\ref{tab:fi_results} gives the $95\%$ confidence interval that applies to each percentage cell in the same column, computed following Leveugle's statistical fault-injection methodology~\cite{Leveugle_2009} with the conservative $p = 0.5$ worst-case binomial assumption, as in recent studies~\cite{Rogenmoser_2025}.

\subsection{Baseline Detection}

The Baseline detection rows of Table~\ref{tab:fi_results} summarise what the existing verification setup observes. Failed-run rates exceed $97\%$ for the full design and the \gls{GNN} under both campaigns, and for the Preprocessing, Postprocessing, and Belle2Link \gls{MAC} sub-components under fixed-rate injection. The Postprocessing Stage and Belle2Link \gls{MAC} retain meaningful pass rates and longer \gls{MTTF} under proportional-rate injection because of their smaller signal share, which means fewer faults per run. Under fixed-rate they drop to near-zero pass, confirming that the proportional-rate pass rates were a low-exposure artefact.

Deadlocks are observed exclusively in the \gls{GNN} Dataflow Accelerator ($63.58\%$ under proportional-rate and $76.98\%$ under fixed-rate injection), since it is the only block currently covered by the \gls{HLS}-generated watchdog. The Chisel-based sub-components (Preprocessing, Postprocessing, and Belle2Link \gls{MAC}) have no dedicated deadlock detection. Their failures surface either as packet-integrity violations at the design output or as simulation timeouts, both detected only after long propagation latency.

\subsection{Inter-stage Liveness Monitoring}

The Inter-stage liveness monitoring rows of Table~\ref{tab:fi_results} reclassify the same simulation runs using the four cocotb transaction monitors at the \gls{AXI}-Stream interfaces. With this approach, each deadlock is attributed to a specific stage: \texttt{Pre.\ header}, \texttt{GNN DF output}, \texttt{Post.\ CPS}, or \texttt{Post.\ GDL}. The timeout class drops below $0.5\%$ on the Preprocessing and GNN Dataflow Accelerator targets, confirming that most baseline timeouts were deadlocks that the existing setup could not catch directly. \gls{MTTF} deviations range from $38.5$ to $78.7\%$ across the Preprocessing, GNN Dataflow Accelerator, and Postprocessing sub-components, with Preprocessing under fixed-rate injection at the upper end, as shown in the MTTF deviation row.

The HLS watchdog row drops below $1\%$ in every inter-stage column because the inter-stage monitors take priority and tag the same GNN deadlocks first. A stall in Postprocessing back-pressures the GNN-to-Postprocessing interface and freezes the \gls{GNN}'s output stream. As a result, $4.13\%$ of Postprocessing-injected runs under proportional-rate and $33.14\%$ under fixed-rate are classified as \texttt{GNN DF output} deadlocks, even though the fault originated in Postprocessing.

\subsection{Discussion}

\glspl{SDC} are naturally observed at the design output: intermediate errors do not affect the measurement as long as the output is correct, whether because no error occurred or because errors were masked along the way. For \glspl{DUE} the opposite holds. A fault that stalls the pipeline at time $t_d$ is only registered as a failure at $t_d + \Delta$, where $\Delta$ is the detection latency. Since the detection latency can not be removed from the \gls{MTTF} calculation, the sooner an error is observed, the more accurate the measurement, especially when multiple faults are injected per run to accelerate the campaign. Output-only observation, as in~\cite{Xu_2021}, therefore inflates \gls{MTTF} estimates for \glspl{DUE}. Inter-stage monitors close this gap, except for the Belle2Link \gls{MAC}, where all monitors are placed before it in the pipeline and $\Delta$ remains identical to the baseline, leaving the \gls{MTTF} deviation at zero.

Beyond closing detection gaps, the inter-stage \gls{MTTF} data identifies hardening priorities for the design. Under proportional-rate injection, the Preprocessing Stage and \gls{GNN} Dataflow Accelerator have the shortest \glspl{MTTF}, making them the highest-priority hardening targets. The Belle2Link \gls{MAC} and Postprocessing Stage have substantially longer \glspl{MTTF} and contribute less to the Full Design's failure rate. The fixed-rate campaign reaches the same conclusion, even though smaller sub-components are injected at fault rates up to 35x higher than in the proportional-rate campaign.

%% file: figures/tab_campaign_summary.tex
\begin{table}[t]
    \centering
    \caption{Per-target injection rate, simulation runs, and total actual injected faults in the two campaigns.}
    \label{tab:campaign_summary}
    \setlength{\tabcolsep}{2.5pt}
    \scriptsize
    \begin{tabular*}{\columnwidth}{@{\extracolsep{\fill}}lrrr}
        \toprule
        & \textbf{\scriptsize Runs} & \textbf{\scriptsize Rate (ns/fault)} & \textbf{\scriptsize Faults inj.} \\
        \midrule
        \multicolumn{4}{l}{\textit{Proportional-rate campaign}} \\
        Full Design                     & 5\,908 &  1\,846 & 332\,717 \\
        Preprocessing Stage             & 3\,196 &  8\,141 &  63\,053 \\
        GNN Dataflow Accelerator        & 3\,630 &  3\,357 & 133\,173 \\
        Postprocessing Stage            & 7\,123 & 64\,885 &  26\,254 \\
        Belle2Link MAC                  & 5\,973 &  9\,515 &  97\,712 \\
        \midrule
        \multicolumn{4}{l}{\textit{Fixed-rate campaign}} \\
        Full Design                     & 5\,908 &  1\,846 & 332\,717 \\
        Preprocessing Stage             & 2\,872 &  1\,846 & 217\,092 \\
        GNN Dataflow Accelerator        & 2\,711 &  1\,846 & 145\,855 \\
        Postprocessing Stage            & 3\,730 &  1\,846 & 286\,867 \\
        Belle2Link MAC                  & 2\,863 &  1\,846 & 140\,117 \\
        \bottomrule
    \end{tabular*}
\end{table}

%% file: figures/tab_fi_results_T.tex
\begin{table*}[t]
    \centering
    \caption{Fault injection results per sub-component, under baseline and inter-stage-liveness detection, for both injection campaigns.}
    \label{tab:fi_results}
    \setlength{\tabcolsep}{4pt}
    \small
    \begin{tabular*}{\textwidth}{@{\extracolsep{\fill}}lrrrrr@{\hspace{10pt}\vrule\hspace{4pt}}rrrrr}
        \toprule
        & \multicolumn{5}{c@{\hspace{10pt}\vrule\hspace{4pt}}}{\textbf{\textit{Proportional-rate campaign}}} & \multicolumn{5}{c}{\textbf{\textit{Fixed-rate campaign}}} \\
        \cmidrule(lr{10pt}){2-6}\cmidrule(l{10pt}r){7-11}
        & \textbf{\scriptsize Full} & \textbf{\scriptsize Pre.} & \textbf{\scriptsize GNN DF} & \textbf{\scriptsize Post.} & \textbf{\scriptsize B2L}
        & \textbf{\scriptsize Full} & \textbf{\scriptsize Pre.} & \textbf{\scriptsize GNN DF} & \textbf{\scriptsize Post.} & \textbf{\scriptsize B2L} \\
        \midrule
        Runs ($n$)               & 5\,908 & 3\,196 & 3\,630 & 7\,123 & 5\,973 & 5\,908 & 2\,872 & 2\,711 & 3\,730 & 2\,863 \\
        Passed (\%)              &   0.02 &   1.16 &   2.23 &  52.93 &  21.56 &   0.02 &   0.03 &   0.07 &   0.03 &   0.21 \\
        \midrule
        \multicolumn{11}{l}{\textbf{\textit{Baseline detection}}} \\
        HLS watchdog (\%)        &  46.07 &   0.00 &  63.58 &   0.00 &   0.00 &  46.07 &   0.00 &  76.98 &   0.00 &   0.00 \\
        Packet integrity (\%)    &  42.57 &  68.87 &  22.89 &  29.07 &  67.09 &  42.57 &  70.68 &  16.31 &  72.89 &  90.08 \\
        Sim timeout (\%)         &  11.34 &  29.97 &  11.30 &  18.00 &  11.35 &  11.34 &  29.29 &   6.64 &  27.08 &   9.71 \\
        MTTF ($\mu$s)            &  110.4 &  164.3 &  129.7 &  509.2 &  236.4 &  110.4 &  142.0 &  102.6 &  147.5 &  108.8 \\
        \midrule
        \multicolumn{11}{l}{\textbf{\textit{Inter-stage liveness monitoring}}} \\
        HLS watchdog (\%)        &   0.37 &   0.00 &   0.33 &   0.00 &   0.00 &   0.37 &   0.00 &   0.66 &   0.00 &   0.00 \\
        Packet integrity (\%)    &  36.11 &  60.36 &  20.83 &  28.36 &  67.09 &  36.11 &  43.21 &  14.17 &  45.58 &  90.08 \\
        Sim timeout (\%)         &   0.00 &   0.19 &   0.41 &  12.28 &  11.35 &   0.00 &   0.00 &   0.00 &   0.56 &   9.71 \\
        Pre.\ header (\%)        &  20.43 &  38.26 &   0.00 &   0.00 &   0.00 &  20.43 &  56.58 &   0.00 &   0.00 &   0.00 \\
        GNN DF output (\%)       &  42.55 &   0.03 &  76.20 &   4.13 &   0.00 &  42.55 &   0.07 &  85.10 &  33.14 &   0.00 \\
        Post.\ CPS (\%)          &   0.20 &   0.00 &   0.00 &   0.94 &   0.00 &   0.20 &   0.11 &   0.00 &   8.31 &   0.00 \\
        Post.\ GDL (\%)          &   0.32 &   0.00 &   0.00 &   1.36 &   0.00 &   0.32 &   0.00 &   0.00 &  12.38 &   0.00 \\
        MTTF ($\mu$s)            &   44.1 &   78.3 &   79.7 &  486.4 &  236.4 &   44.1 &   30.3 &   46.8 &   59.5 &  108.8 \\
        \midrule
        \textbf{MTTF deviation (\%)} & \textbf{60.0} & \textbf{52.3} & \textbf{38.5} & \textbf{4.5} & \textbf{0.0} & \textbf{60.0} & \textbf{78.7} & \textbf{54.4} & \textbf{59.7} & \textbf{0.0} \\
        Margin of error (95\%)   & $\pm1.27$ & $\pm1.73$ & $\pm1.63$ & $\pm1.16$ & $\pm1.27$ & $\pm1.27$ & $\pm1.83$ & $\pm1.88$ & $\pm1.60$ & $\pm1.83$ \\
        \bottomrule
        \addlinespace[1pt]
        \multicolumn{11}{l}{\scriptsize Full = Full Design; Pre.\ = Preprocessing Stage; GNN DF = GNN Dataflow Accelerator; Post.\ = Postprocessing Stage; B2L = Belle2Link MAC.} \\
    \end{tabular*}
\end{table*}

%% file: sections/6_conclusion.tex
\section{Conclusion}
\label{sec:conclusion}

We have presented the first \gls{RTL} fault-injection study of a deployed \gls{L1} hardware \gls{NN}-trigger, targeting GNN-ETM in the Belle~II trigger system. Through two campaigns, a proportional-rate campaign that reflects deployment-realistic exposure and a fixed-rate campaign that equalises exposure across sub-components, we injected $1\,442\,840$ \glspl{SEU} across the $211\,245$ signals of the full design and its four sub-components.

We characterised three failure modes most consequential to a real-time trigger pipeline: deadlocks, timeouts, and packet-integrity violations. The existing verification infrastructure shows a monitoring asymmetry, as deadlocks are observable only inside the \gls{HLS}-based \gls{GNN} Dataflow Accelerator, while failures in the Chisel-based sub-components surface only as packet-integrity violations or simulation timeouts.

We proposed inter-stage liveness monitoring as a more accurate alternative to output-only observation. \gls{MTTF} estimates from the two approaches differ by up to 78.7\%, exposing the bias of measuring failures only at the design output. The Belle2Link \gls{MAC} is not covered by this gain, since all four monitors sit upstream of it. A fifth monitor at the b2link output would close this gap, though with longer detection latency since b2link packets arrive less often. These monitors are not part of the accelerator pipeline, so they could be implemented as a hardware IP running in parallel with the accelerator without affecting the latency budget, and a reset on detected deadlock would restore normal operation most of the time. The per-stage attribution under proportional-rate injection identifies the \gls{GNN} Dataflow Accelerator and the Preprocessing Stage as the highest-priority hardening targets, which will be the focus of our follow-up work.